\documentclass[fleqn,usenatbib]{mnras}
\usepackage{newtxtext,newtxmath}
\usepackage[normalem]{ulem}
\usepackage{tabularx}
\newcolumntype{Y}{>{\raggedright\arraybackslash}X}  
\usepackage[T1]{fontenc}

\DeclareRobustCommand{\VAN}[3]{#2}
\let\VANthebibliography\thebibliography
\def\thebibliography{\DeclareRobustCommand{\VAN}[3]{##3}\VANthebibliography}

\usepackage{graphicx}	
\usepackage{amsmath}	
\usepackage{microtype}
\usepackage{xcolor}
\usepackage{xspace}
\def\HI{\ion{H}{I}~}
\def\Hi{\ion{H}{I}}
\def\HII{\ion{H}{II}~}

\def\xb{\bar{x}_{\rm \ion{H}{I}}}

\def\xa{\bar{x}_{\rm \ion{H}{II}}}

\def\lya{Ly$\alpha$\xspace}

\usepackage{lipsum} 
\usepackage{comment}

\usepackage{tikz,xcolor,hyperref}
\definecolor{lime}{HTML}{A6CE39}
\DeclareRobustCommand{\orcidicon}{%
	\begin{tikzpicture}[baseline=-0.8ex]
	\draw[lime, fill=lime] (0,0) 
	circle [radius=0.16] 
	node[white] {{\fontfamily{qag}\selectfont \tiny ID}};
	\draw[white, fill=white] (-0.0625,0.095) 
	circle [radius=0.007];
	\end{tikzpicture}
	\hspace{-2mm}
}

\foreach \x in {A, ..., Z}{%
	\expandafter\xdef\csname orcid\x\endcsname{\noexpand\href{https://orcid.org/\csname orcidauthor\x\endcsname}{\noexpand\orcidicon}}
}

\title[RMFP-BSD from EoR]{Radial mean free path bubble size distribution (RMFP-BSD) from the Epoch of Reionization} 

\author[I. Georgiev et al.]{
Ivelin Georgiev$^{1,2}$\thanks{E-mail: igeorgiev@mpa-garching.mpg.de}\orcidA,
Robert Crawford$^{3}$, 
Saleem Zaroubi$^{3,2}$\orcidB,
Raghunath Ghara$^{4}$\orcidC,
\newauthor
Abinash Kumar Shaw$^{1,2}$\orcidD,
Garrelt Mellema$^{5}$\orcidE,
Benedetta Ciardi$^{1}$\orcidF,
L. V. E. Koopmans$^{3}$\orcidG,
\newauthor
Anshuman Acharya$^{6}$\orcidH,
Sambit K. Giri$^{5}$\orcidI,
Ilian T. Iliev$^{7}$\orcidJ,
and Qing-bo Ma$^{8,9}$\orcidK
\\
$^{1}$Max-Planck-Institut für Astrophysik, Karl-Schwarzschild-Straße 1, 85748 Garching, Germany\\
$^{2}$ARCO (Astrophysics Research Center), Department of Natural Sciences, The Open University of Israel, 1 University Road, PO Box 808, Ra'anana 4353701, Israel\\
$^{3}$Kapteyn Astronomical Institute, University of Groningen, PO Box 800, 9700 AV Groningen, The Netherlands\\
$^{4}$Department of Physics, Indian Institute of Technology Kharagpur, Kharagpur, West Bengal 721302, India\\
$^{5}$Department of Astronomy and Oskar Klein Centre, AlbaNova, Stockholm University, SE-10691 Stockholm, Sweden\\
$^{6}$Berkeley Center for Cosmological Physics (BCCP), University of California, Berkeley, CA 94720, United States\\
$^{7}$Astronomy Centre, Department of Physics \& Astronomy, Pevensey III Building, University of Sussex, Falmer, Brighton, BN1 9QH, United Kingdom\\
$^{8}$School of Physics and Electronic Science, Guizhou Normal University, Guiyang 550001, PR China\\
$^{9}$Guizhou Provincial Key Laboratory of Radio Astronomy and Data Processing, Guizhou Normal University, Guiyang 550001, PR China
}

\date{Submitted to MNRAS. Received YYY; in original form ZZZ}

\pubyear{2026}

\begin{document}
\label{firstpage}
\pagerange{\pageref{firstpage}--\pageref{lastpage}}
\maketitle


\begin{abstract}
Among the many statistics used to characterise ionised regions during the Epoch of Reionization (EoR), we introduce one which, while closely related to the mean free path bubble size distribution method from \citet{Mesinger2007}, is analytically tractable and substantially less dispersive than that method. The radial mean free path (\texttt{rmfp}) method casts six rays along the Cartesian axes in a 3D volume from a random origin to a neutral hydrogen island. This allows us to better quantify the role of overlap during the EoR, identify isolated or interconnected clusters in the simulation, and study the characteristic peak in the ionisation power spectrum, which is important for 21-cm studies of the intergalactic medium. Moreover, we find the method informative about the properties of neutral $\Hi$ islands during the final phases of the EoR. We also show that applying both methods in tandem can further improve our insight into the morphology of reionization. 
\end{abstract}



\section{Introduction}
\label{sec:intro}
The Epoch of Reionization (EoR) is a key period within the first billion years of the Universe, when the light from the first stars, galaxies, and accreting black holes reionised the neutral intergalactic medium (IGM) \citep{Zaroubi2013}. Studies of the ionisation morphology can yield critical insight into the first ionising sources and are actively being explored using \lya damping wings \citep{Mason2026}, \lya emitting galaxies \citep{Hayes2023}, as well as quasar spectra \citep{Bosman2022}. However, it is the 21-cm signal from the neutral hydrogen in the IGM during the EoR that offers the most direct three-dimensional probe of this process \citep{Giri2018}. Hence, a multitude of methods have been applied to characterise the properties of ionised regions in simulations such as the Friends of Friends \citep{Iliev2006a}, Minkowski functionals \citep{Gleser2006}, Spherical-Average \citep{Zahn2007}, Granulometry \citep{Kakiichi2017}, Opening transforms \citep{Busch2020}, Watershed \citep{Lin2016}, and Betti numbers \citep{giri2021measuring}. Comparative studies such as those of \citet{Lin2016}, \citet{Kakiichi2017}, and \citet{Giri2018} have examined these methods, each favouring a different approach. Here we adopt the mean free path bubble size distribution method (MFP-BSD, hereafter \texttt{mfp}) \citep{Mesinger2007}, which \citet{Giri2018} found to perform best at tracking the progress of reionization and separating different reionization scenarios. In this method, random rays are drawn within an ionised region and traced up to a neutral island. As a result, it samples both very short and very long lines of sight, which broadens the distribution. In this work, we introduce a new method inspired by the \texttt{mfp} that reduces this smearing and, in doing so, introduces a novel summary statistic to the community. We briefly introduce the theory in Sec.~\ref{sec:theory}, applications in Sec.~\ref{sec:results}, and conclude in Sec.~\ref{sec::conclusions}.
\section{Method}
\label{sec:theory}
In this section, we briefly summarise the theory behind the radial mean free path bubble size distribution method (\texttt{rmfp}) and how it differs from the \texttt{mfp} method. In general \citep[see][]{Solomon1978}, the probability density of the radial position $t$ of a point drawn uniformly inside a sphere of radius $R$ is defined as 
\begin{equation}
\label{eq:sphere_density}
        \frac{\mathrm{d}P(t)}{\mathrm{d}t} = \frac{3 t^{2}}{R^{3}},\,\mathrm{where}\,t=\sqrt{x^{2}_{0}+y^{2}_{0}+z^{2}_{0}} 
\end{equation}
and $x_0$, $y_0$, and $z_0$ are the origin points in Cartesian space. In the \texttt{rmfp} methodology, once we choose a random point within the sphere, we sample rays aligned with an orthonormal basis (i.e., the $x$, $y$, and $z$ axes). Rays are launched in positive and negative directions from the starting point, and we calculate the mean of the two lengths along each axis (e.g. $r_{x} = (x_{\mathrm{+}}+x_{\mathrm{-}})/2$). We can then calculate the ray length to the edge as the quadrature mean of the three axis lengths $r_{e} =\sqrt{R^{2}-\frac{2}{3}t^{2}}$, so that $r_{e}=R$ when the sampling point coincides with the centre and $r_{e}=R/\sqrt3$ when it lies on the surface. We can define the logarithmically weighted distribution as
\begin{equation}
\label{eq:rmfp-bsd}
     r\frac{\mathrm{d}P_{\mathrm{rmfp}}(r)}{\mathrm{d}r} = \frac{9r^2}{2R^{2}}\sqrt{\frac{3}{2}\Big(1-\frac{r^{2}}{R^{2}}\Big)},\,\mathrm{for}\, r \in [R/\sqrt{3},R].
\end{equation}
We apply extrema and moment analysis to estimate the peak of the distribution at $\bar{r}_{\mathrm{rmfp}} = \sqrt{2/3}R$, the mean value at $\mu_{\mathrm{rmfp}} = 0.783 R$ with a standard deviation of $\sigma_{\mathrm{rmfp}} = 0.113 R$\footnote{The same logarithmic weighting is applied in Eq.~\eqref{eq:mfp-bsd}, and all moments quoted here ($\bar{r}$, $\mu$, $\tilde{x}$, $\sigma$) refer to the normalised curve $r\,\mathrm{d}P/\mathrm{d}r$, which is the quantity plotted throughout this work.}. We can also theoretically estimate the median to be $\tilde{x}_{\mathrm{rmfp}} = 0.785 R$.

For the sake of clarity in comparison, we can also consider the classic \texttt{mfp} distribution, and following \citet{Lin2016} we define radius $r$ to the edge of the bubble, with a ray emitted at some random angle $\theta$ such that $r_e = \sqrt{R^{2}- t^2 \sin^2\theta} - t\cos\theta$, ranging from 0 to 2$R$. Integrating across all possible radial positions and initialisation angles, we can express
\begin{equation}
\label{eq:mfp-bsd}
     r\frac{dP_{\mathrm{mfp}}(r)}{dr} = \frac{3}{16} \frac{r}{R} \Big( 4-\frac{r^2}{R^2}\Big ) ,\,\mathrm{for}\, r \in [0,2R].
\end{equation}
Hence, we can infer that the peak of the distribution is at $\bar{r}_{\mathrm{mfp}} = 2R/\sqrt3$, its mean $\mu_{\mathrm{mfp}} = 16R/15$, whilst the standard deviation is $\sigma_{\mathrm{mfp}} = 2\sqrt{11}R/15\approx 0.442 R$ and the median $\tilde{x}_{\mathrm{mfp}} = 1.082 R$. We note that the ratio of the two modes is exactly $1/\sqrt{2}$, close to that of the medians. We adopt the median throughout this work, since it is insensitive to the binning and smoothing of the measured distribution. Its uncertainty follows directly from the quantiles, whereas the mode is sensitive to both.
\begin{figure}
	\includegraphics[width=1.0\columnwidth]{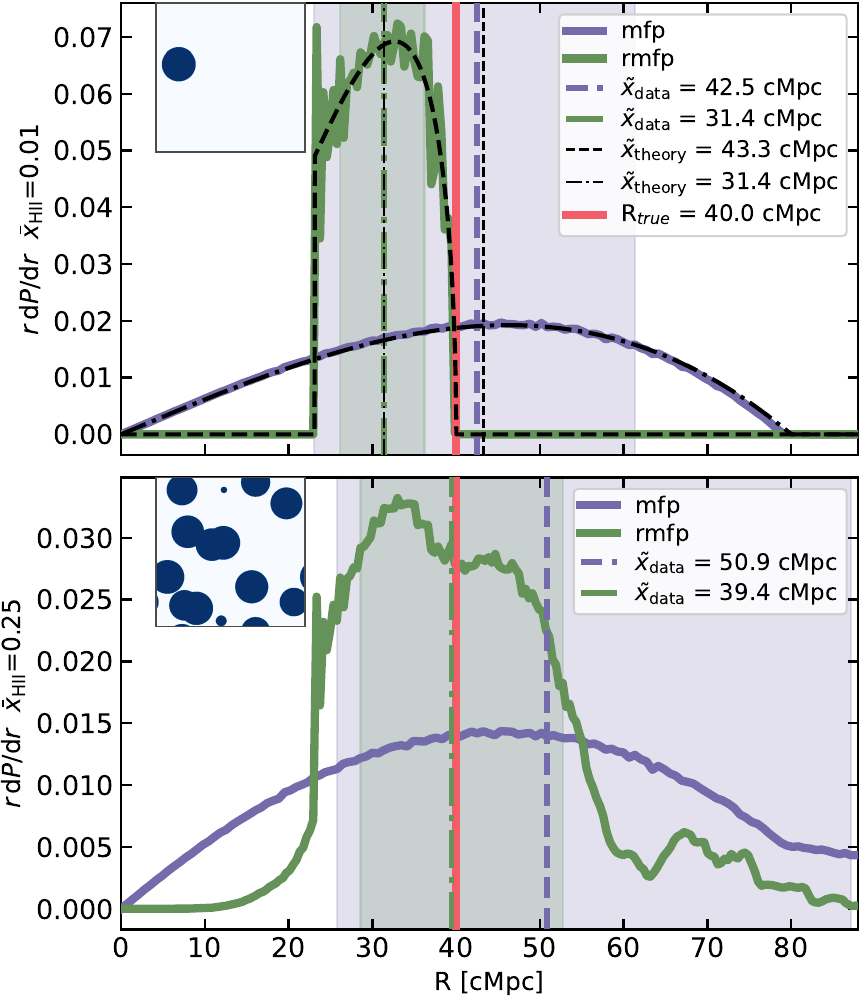}
    \caption{Comparison between the \texttt{mfp} and \texttt{rmfp} for a case with one bubble ($R=40$ cMpc, shown as a thick vertical line) in the upper panel and fifty overlapping bubbles of the same size in the bottom panel. The thin dashed lines in the upper panel represent the theoretical PDFs estimated from equations~\ref{eq:rmfp-bsd} and \ref{eq:mfp-bsd}, while the thin dashed vertical lines mark the median of each distribution. The distributions are also derived numerically by populating a box with a volume of (349 cMpc)$^{3}$ (seen in the small square panel at the top left of every panel), with the medians estimated from the data shown as thick vertical lines. The fluctuations in the \texttt{rmfp} distribution result from resolution effects when sampling the ionisation field.}
    \label{fig:simple_R}
\end{figure}
\begin{figure*}
    \centering
    \includegraphics[width=1.\linewidth]{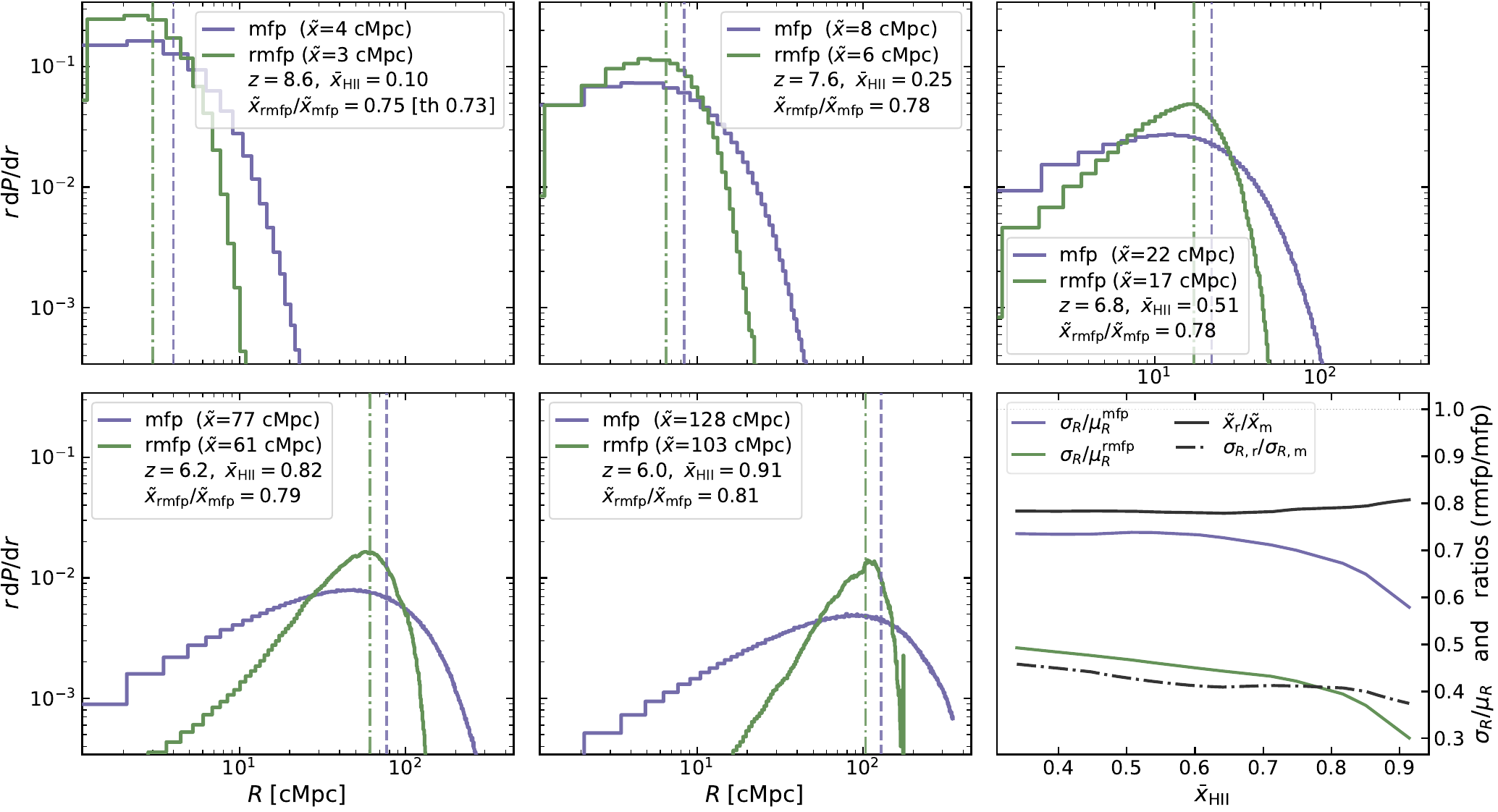}
    \caption{Distribution of mean free paths for the \texttt{pLLS} model during the EoR using the \texttt{mfp} and \texttt{rmfp} methods; their median values are shown with a dashed and dashed-dotted line, respectively. The final rightmost panel shows the evolution of the coefficient of variation $\sigma/\mu$ during the EoR, as well as the ratios of the methods' median $\tilde{x}$ and dispersion $\sigma$.}
    \label{fig:bsd}
\end{figure*}

We can contextualise the probabilities by directly examining the two distributions numerically in Fig.~\ref{fig:simple_R}. The top panel shows a single bubble of radius $R=40$ cMpc, the bottom panel shows fifty such bubbles randomly distributed in a box of side length $L = 349$ cMpc, resulting in a binary ionisation field. We choose to show the median ($\tilde{x}$) of each distribution, as it is less sensitive to outliers and is used in the following sections. We show the median, estimated theoretically from equations \eqref{eq:rmfp-bsd} \& \eqref{eq:mfp-bsd}, as thin vertical lines, as well as directly from the PDFs in the numerical study (thick vertical lines). The \texttt{mfp} method (matching the analytic prediction) recovers the true radius to within 8 per cent. The flatter peak of the distribution results from visualising the distribution per logarithmic interval in $r$, which suppresses the small-$r$ end where $\mathrm{d}P/\mathrm{d}r$ is largest and shifts the maximum to $2R/\sqrt{3}$. In contrast, the \texttt{rmfp} method underestimates the true radius, but with a much smaller spread, and its median falls within the uncertainty of the \texttt{mfp} method. The underestimation is understood by examining Eq.~\eqref{eq:sphere_density}. We infer that because most of the sphere's volume lies near the edge, most rays underestimate the true radius. However, the tighter spread around the \texttt{rmfp} is more informative of the bubble internal structure. By design, the method cannot sample radii below $R/\sqrt{3}$ or above $R$, and therefore, any changes in the width of the spread will be informative either of separate smaller ionised regions or of larger, overlapping structures. 

We emulate this numerically in the bottom panel of Fig.~\ref{fig:simple_R} by distributing fifty spheres, each with a radius of $R= 40$ cMpc, such that the globally averaged ionisation fraction is $\xa \approx 0.25$ and 21 per cent of the total ionised volume lies in overlap regions. As expected, the true size is within the \texttt{mfp} spread, yet the distribution is now too broad to localise it. The shape of the distribution has now become considerably more smeared as the method is more sensitive to the longest lines of sight. We can naïvely think of the uncertainty as a sum of all possible contributions from that of a single bubble, the overlapping structures, or the presence of larger bubbles. Hence, the increased spread will only rise and become less informative as the ionisation topology considered becomes more complex. 

In contrast, the \texttt{rmfp} distribution is now slightly bimodal, with one peak near the single-bubble case and the other slightly above the true value. Therefore, the \texttt{rmfp} still retains knowledge of the original bubble size, but now the mode of the distribution and spread contain additional information on the overlapping structure. The box in the top left of the figure represents this, as we see ``pear-shaped'' ionised regions (in this slice), which can be interpreted either as two identical regions or as one slightly smaller than the sum of its parts. This is reflected by the fact that the median of the distribution now matches the true value of a single bubble, which allows us to infer an effective radius of $R_{\mathrm{overlap}} = \tilde{x}_{\mathrm{data}} /0.785 \approx 50$ cMpc, corresponding to an overlapped bubble structure which can neatly fit within the showcased slice. 

In summary, both distributions are offset from the true radius by a known factor, placing the \texttt{rmfp} median below the \texttt{mfp} one, though still within its $1\sigma$ uncertainty, and this separation narrows once overlap begins. Since the offset is predictable, it can therefore be accounted for when studying the EoR with this metric. Moreover, the spread in the \texttt{rmfp} distribution is narrower than that of the \texttt{mfp}, and is therefore more informative about the role and evolution of overlapping structures during the EoR. We note that since the \texttt{rmfp} method samples volume-weighted points within \HII regions, it is therefore complementary to, rather than a replacement for, sight-line statistics measured from fixed source positions.
\begin{figure}
	\includegraphics[width=1.0\columnwidth]{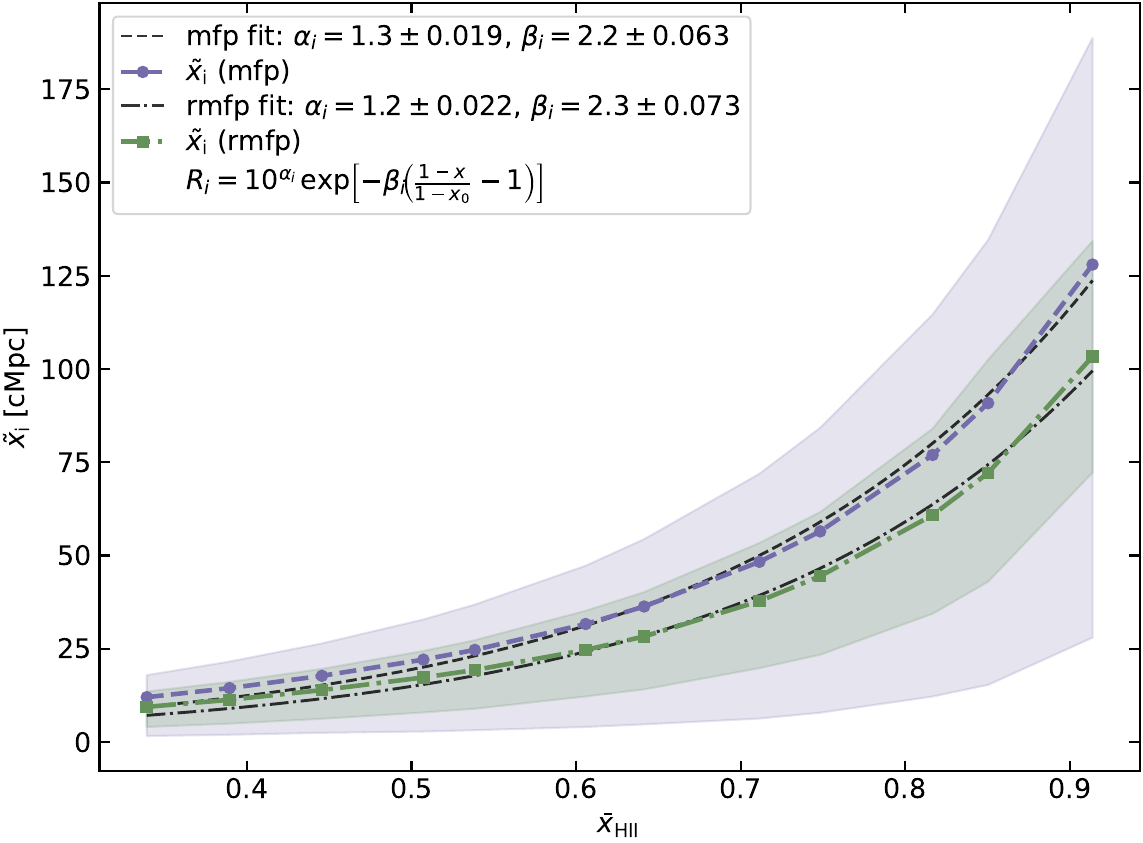}
	\includegraphics[width=1.0\columnwidth]{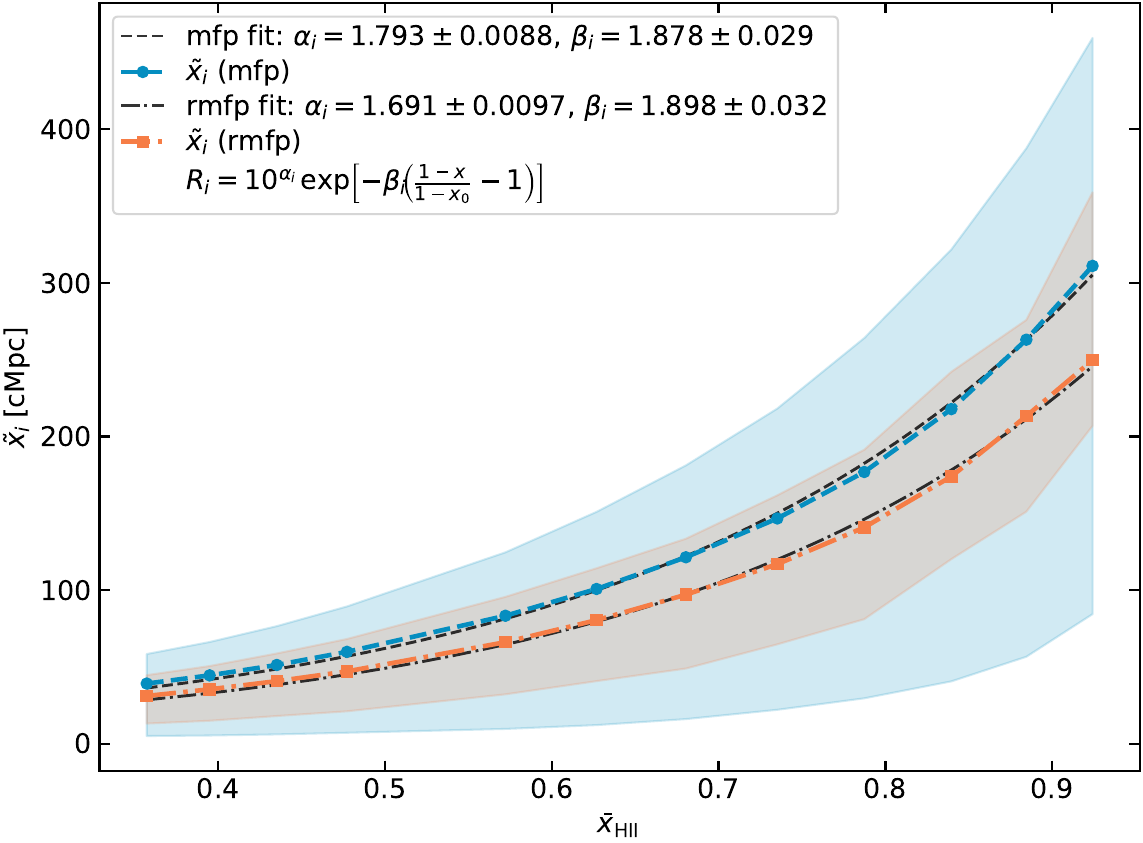}
    \caption{Evolution of the median values and their $1 \sigma$ uncertainty of the sizes of \HII regions from the \texttt{pLLS} and \texttt{C0} models (upper and lower panels, respectively), produced by the \texttt{mfp} and \texttt{rmfp} methods. An analytic fit of the form $R_{i} = 10^{\alpha_{i}}\exp{\big[-\beta_{i}((1-\xa)/(1-x_0)-1)\big]}$ with a pivot at $x_0 = 0.5$ is fitted to the data (thin grey lines), and the free parameters are shown in the legend for a cleaner comparison between the methods.}
    \label{fig:med_growth}
\end{figure}
\section{Results}
\label{sec:results}
We utilise the 3D radiative transfer EoR simulation with a comoving volume of (349 cMpc)$^3$, gridded onto $250^3$ cells, from \citet{Georgiev2025}, which we will refer to as \texttt{pLLS}. Ionising sources are assigned a constant efficiency parameter $\zeta = 50$, while the opacity of the intergalactic medium is set by a position-dependent distribution of Lyman limit systems, following an approach similar to that of \citet{Shukla2016}.

We apply both methods to the \texttt{pLLS} simulation, initiating $10^{6}$ random rays at each redshift output, and present the distributions inferred from the ionised regions in Fig.~\ref{fig:bsd}. In the legend, we include the ratio of the median from each distribution and its analytic prediction of $\tilde{x}_{\mathrm{rmfp}}/\tilde{x}_{\mathrm{mfp}} = 0.785R/1.082R \approx0.725$ for a single sphere from Sec.~\ref{sec:theory}. The predicted value is broadly consistent early on in the EoR ($\xa \approx 0.10$), seen in the top panel of the first column, where the median of the \texttt{mfp} distribution is larger than that of the \texttt{rmfp}. Naturally, in this early regime percolation has yet to take place \citep[$\xa \approx 0.15$, see][]{Furlanetto2016}, and the ionised morphology is still driven largely by ``\HII bubbles'' growing around their local overdensities. Conversely, in the second panel of the first row, the ionisation morphology is characterised by a large number of overlapping ionised regions. The ratio between the medians increases slightly in the remaining panels, as the role of overlapping structures becomes ever more pronounced. The standard deviation of each distribution is also shown in the bottom-right panel of Fig.~\ref{fig:bsd}, where we plot the evolution of the coefficient of variation $\sigma/\mu$, the ratio of the standard deviations between the two methods, and the ratio of the median values. For the \texttt{mfp} method, the coefficient of variation indicates a scatter of around 70 per cent around the mean value, which remains more or less constant through the EoR, decreasing slightly at the final stages. For the \texttt{rmfp} method, the coefficient of variation is lower by roughly a factor of two, and decreases as reionization proceeds.

As discussed in Sec.~\ref{sec:theory}, the predicted spread in the \texttt{mfp} method is larger than that of the \texttt{rmfp}, and while it probes line-of-sight variance well, it is less informative about overlapping structures. Furthermore, the \texttt{rmfp} spread might more accurately capture the variance of ionised regions later in the EoR. Because of the lower limit placed on the sampling of the rays in Eq.~\eqref{eq:rmfp-bsd}, values lower than $\tilde{x}/\sqrt{3}$ 
seen in the bottom panel of Fig.~\ref{fig:bsd} are likely most indicative of populations of small ionised regions embedded in the neutral medium that have yet to merge onto the largest clusters.
\subsection{Global evolution of the sizes of ionised regions}
The similar evolution of the ratio of the medians, presented in the first five panels of Fig.~\ref{fig:bsd} (and explicitly shown in the final panel against the ionisation fraction), is a proof-of-concept that the \texttt{rmfp} method traces the growth of the same average population of ionised regions as the \texttt{mfp} method. To test this claim, we examine the median distribution (and its uncertainty) for both methods, shown in Fig.~\ref{fig:med_growth}. We compare the \texttt{pLLS} simulation with the \texttt{C0} simulation, identical in all respects except the ionising-photon absorption mechanism. Recombinations are turned off by setting the clumping factor to zero, and the maximum distance photons are allowed to travel is boosted up to $R_{\max} = 100$ cMpc. As a result, the morphology of the ionised regions differs significantly. We fit the growth of the median values using the exponential model of bubble growth applied in eq.~24 of Georgiev et al. (2026a, submitted). The model is defined by two parameters. The $\alpha_{i}$ parameter captures the average size of ionised regions at the midpoint of reionization, where the model is anchored. The $\beta_{i}$ parameter in the exponent folds source and sink properties into a generic evolution parameter related to the rate of the EoR. We note that the fit closely tracks the growth of the ionised regions. While the theory predicts $\tilde x_{\rm rmfp}/\tilde x_{\rm mfp}\approx0.725$ (see rightmost bottom panel of Fig.~\ref{fig:bsd}), both the \texttt{pLLS} and \texttt{C0} ratios produce the same value of $\tilde{x}_{\mathrm{rmfp}}/\tilde{x}_{\mathrm{mfp}} \approx 0.794$, close to the single bubble case though slightly boosted due to overlap. Moreover, the uncertainty ascribed to the \texttt{rmfp} method contains the median value of the \texttt{mfp} method. Furthermore, both distributions describe the evolution of the same bubble population, as the evolution parameter $\beta_i$ is essentially equal between the two methods within $1\sigma$.

For the \texttt{pLLS} model, the average bubble size driving reionization during the EoR is around 15-20 cMpc, whilst for \texttt{C0} the values are in the range of 50-65 cMpc. This is consistent with expectations, as in the former model the growth of ionised structures is impeded by recombinations in the ionised medium and by small-scale absorbers, effects that have been removed in the latter model. This also impacts the evolution parameter $\beta_i$, which is roughly 20 per cent higher for the \texttt{pLLS} model.
\begin{figure*}
    \centering
    \includegraphics[width=.99\linewidth]{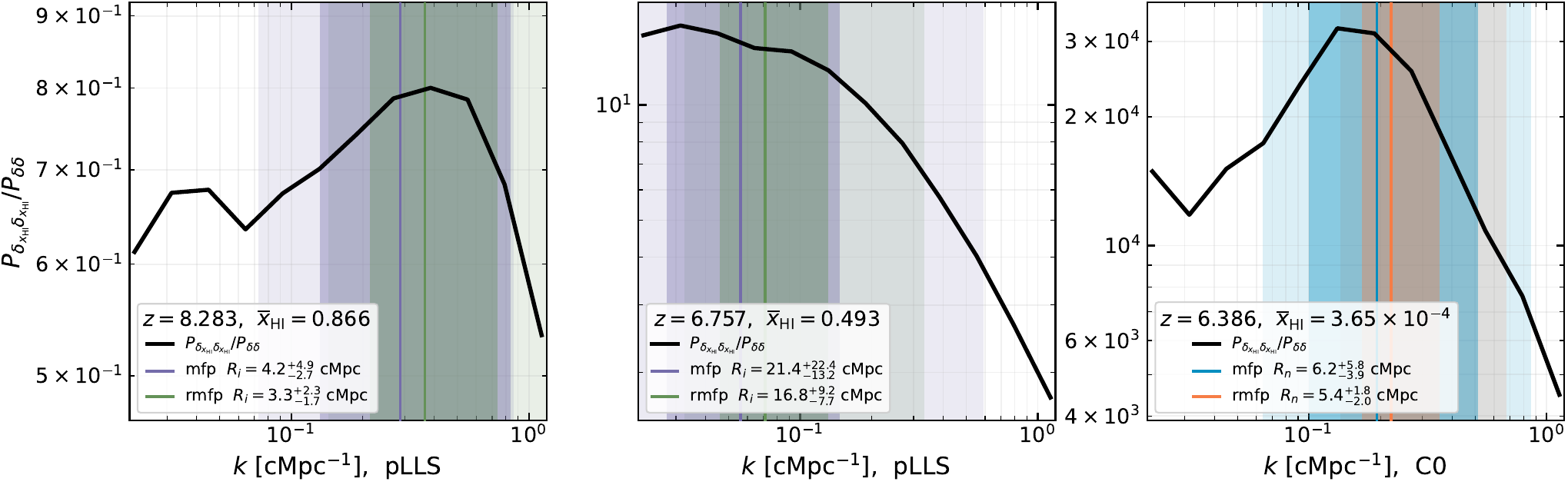}
    \caption{Ratio of the neutral and density autocorrelation power spectra at the beginning and midpoint of the EoR for the \texttt{pLLS} model (first two panels) as well as the very final phases of the EoR in the rightmost panel for the \texttt{C0} model. For each redshift, the median value (and its spread) is plotted at $k = 4/(\pi R)$, following eq.~16 of Georgiev et al. (2026a, submitted).}
    \label{fig:ps}
\end{figure*}
\begin{figure}
	\includegraphics[width=1.0\columnwidth]{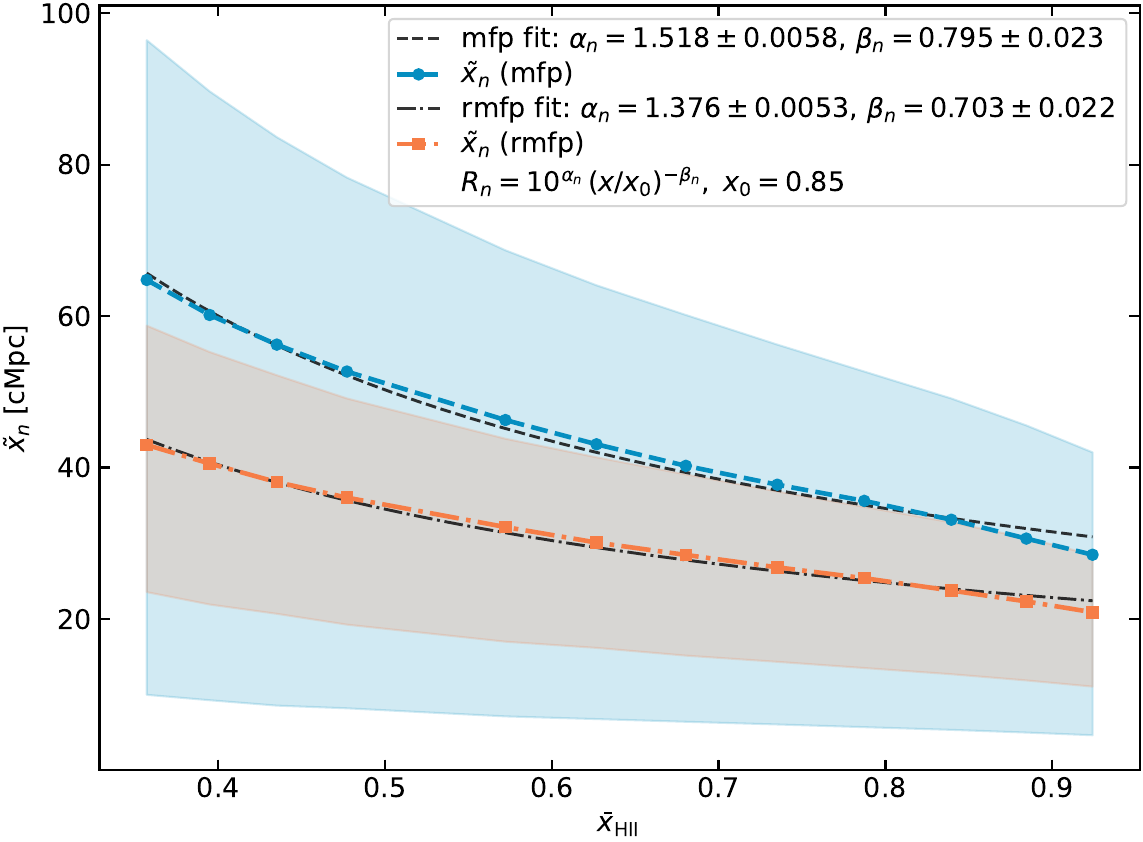}
    \caption{Evolution of the median values and their $1 \sigma$ uncertainty of the sizes of \HI regions from the \texttt{C0} simulation, produced by the \texttt{mfp} and \texttt{rmfp} methods, shown in blue and orange, respectively. An analytic fit of the form $R_{n} = 10^{\alpha_{n}}(\xa/x_0)^{-\beta_{n}}$, with a pivot at $x_0 = 0.85$, is fitted to the data (thin grey lines), and the free parameters are shown in the legend for a cleaner comparison between the methods.} 
    \label{fig:n_med_growth}
\end{figure}
\subsection{Power spectrum of the ionisation morphology}
We further test the application of the method in Fig.~\ref{fig:ps}, where we examine the autocorrelation power spectrum of the fluctuations of the neutral field divided by that of the density fluctuations $P_{\delta_{\mathrm{xHI}},\delta_{\mathrm{xHI}}}/P_{\delta,\delta}$ for the \texttt{pLLS} model. We examine fluctuations in the neutral field, an essential component of the 21-cm power spectrum, to further isolate the role of the ionisation morphology. We study the ratio to account for correlations with the density field and select two strategic epochs. The first panel showcases the power spectrum at the onset of percolation, where the ionisation topology is increasingly driven by ionised regions rather than isolated bubbles, but the role of overlap is still small. Conversely, the middle panel shows the middle of reionization, where fluctuations between the neutral and ionised patches become comparable, and the characteristic size of ionised regions can be compared with the fit prediction in Fig.~\ref{fig:med_growth}. The median sizes and their respective $1\sigma$ and $2\sigma$ spreads at the given epochs are shown in $k$-space using the numeric ratio $4/(\pi R)$ as an estimate. This corresponds to the point where the amplitude of the transfer function of eq. 16 of Georgiev et al. (2026a, submitted) is halved, and both terms in the denominator become comparable. More clearly, it marks the transition point between the scale-independent and scale-dependent regimes associated with the size of the ionised regions. 

While we cannot rely solely on the exact value of this feature, the spread of the distribution will be crucial for inferring the population of ionised regions driving the characteristic scale in the spectrum. We find that both the \texttt{rmfp} and \texttt{mfp} methods (the former is always within the uncertainty of the latter) approximately predict the characteristic peak of the ionised region during the EoR. However, the \texttt{rmfp} method provides a tighter constraint consistent with the measured features, while the spread in the \texttt{mfp} distribution makes it less constraining, with its $1\sigma$ interval covering $\Delta\log_{10} k = 0.79$ against 0.54 for the \texttt{rmfp}. For example, in the leftmost panel at the percolation stage, the \texttt{rmfp} median seems to lie right on the peak of the characteristic feature, with a $1 \sigma$ range covering the region of interest. Conversely, the \texttt{mfp} median is close to the peak feature, yet its uncertainty covers a wide $k$-range ($10^{-1}<k<10^{0}$ cMpc$^{-1}$). This is further exacerbated in the middle panel with the progress of the EoR as the growth and overlap of ionised clusters push the peak to lower wavenumbers and broaden the spread. This is unsurprising, given Fig.~\ref{fig:bsd}, as the \texttt{mfp} method largely relies on lines of sight alone and the median measures the average line-of-sight distance rather than the size of the host ionised region. Thus, the very long lines of sight seen towards the end of the EoR likely push the predicted value higher. Meanwhile, the setup of the \texttt{rmfp} method is found to be most favourable for studying the features of the power spectrum, since it is largely centred on sampling ionised bubbles rather than sight lines, and has a naturally smaller spread. 
\subsection{Constraining neutral islands at the end of reionization}
Lastly, the same analysis of ionised regions readily applies to the neutral morphology. We showcase this in Fig.~\ref{fig:n_med_growth} for the \texttt{C0} case since it has completely reionised by a redshift of 6 (though note the same trends are also observed for the \texttt{pLLS} model, which reaches a value of $\xa = 0.9$ at a redshift of 6). The panel reaffirms the geometric arguments of both sampling methods by showing the median size of neutral regions during the EoR, using the fit in eq.~25 of Georgiev et al. (2026a, submitted). We note the same theoretical ratio, sensitivity to overlap (though in reverse), and a 12 per cent difference in $\beta_{n}$. When applied to study the neutral medium, the \texttt{mfp} method traces single lines of sight and is thus more sensitive to interconnected regions and tunnels connecting the neutral islands, resulting in larger lines of sight. Conversely, the \texttt{rmfp} method traces six rays along the Cartesian axes simultaneously, and when applied to the neutral IGM, it is less sensitive to the tunnels connecting the neutral islands and therefore more sensitive to isolated neutral islands. In the rightmost panel of Fig.~\ref{fig:ps}, we showcase the ratio of the power spectra at $\xb \approx 3.65 \times 10^{-4}$. In this regime, strictly dominated by neutral islands, we see a prominent peak associated with their characteristic size. Notably, both methods constrain the feature, the \texttt{mfp} being slightly more spread out as expected, with its median value falling within the uncertainty of the \texttt{rmfp}. The conclusion is that the \texttt{rmfp} method too can be used as a measure of neutral islands, as in its calculation it can capture some of the anisotropy of the islands and the metric itself is in synergy with how the power spectrum is constructed. Moreover, the \texttt{rmfp} will be less sensitive to outliers such as tunnels, to which the \texttt{mfp} method will be susceptible.
\section{Conclusions}
\label{sec::conclusions}
We introduced the radial mean free path bubble size distribution (\texttt{rmfp}), a novel summary statistic for studying ionisation morphology during the EoR. The method casts six axis-aligned rays from a random point inside an ionised region to the nearest neutral island. In the case of a single bubble, it is analytically tractable, with a distribution bounded to $[R/\sqrt{3},R]$ and a median $\tilde{x}_{\mathrm{rmfp}} = 0.785R$, and is considerably less dispersive than the \texttt{mfp} method. Although the \texttt{rmfp} method is biased low, with $\tilde{x}_{\mathrm{rmfp}}/\tilde{x}_{\mathrm{mfp}} = 0.725$ for a single bubble, it remains within the $1\sigma$ uncertainty of the \texttt{mfp}. We test this assumption by applying both methods to the \texttt{pLLS} simulation in Fig.~\ref{fig:bsd}. The numerical ratio of the medians, shown in the legend of each panel, remains close to the theoretical value, slowly increasing due to the overlap of ionised regions. This suggests that rescaling $\tilde{x}_{\mathrm{rmfp}}$ could correct the \texttt{rmfp} bias, recovering the true size to the same accuracy as the \texttt{mfp} method while retaining the narrower, more informative spread. Focusing on the evolution of the median of each distribution alone in Fig.~\ref{fig:med_growth}, we confirm that each method tracks the same ionised-region population at the same growth rate, and the tighter \texttt{rmfp} spread makes it markedly more informative about overlapping structures. We also apply the method to study the neutral medium at the end of the EoR. We find it is less sensitive to the tunnels connecting neutral islands and therefore better isolates their characteristic size. Lastly, we test both methods by relating their predicted values to the characteristic peak in the ratio of the neutral-fraction and density power spectra in Fig.~\ref{fig:ps}. While both methods are consistent with the characteristic peak in the power spectra, it is the precision of the \texttt{rmfp} method that proves to be more informative. 

In future work, we will compare the \texttt{rmfp} method against the watershed \citep{Lin2016} and granulometry \citep{Kakiichi2017} approaches, which resolve individual regions rather than sampling them. Lastly, the predictability of the offset in the \texttt{rmfp} distribution suggests it could be rescaled onto the true radius. Since this would require a single multiplicative factor, the shift would leave all relative measures unchanged, such as $\sigma/\mu$ and the quantile ratios, and therefore preserve the information carried by the spread. The factor is, however, derived for a single bubble, and in future work we plan to test this calibration across a wider range of ionisation topologies.
\section*{Acknowledgements}
IG, SZ, BC, AKS acknowledge support from the Israel Science Foundation (grant no. 1388/24) and the GIF “German-Israeli Foundation for Scientific Research and Development''. L.V.E.K. acknowledges support from the European Research Council (ERC) under the European Union’s Horizon 2020 research and innovation programme, grant agreement No. 884760 (CoDEX). AA acknowledges financial support from the Simons Foundation for their current position as a ``Learning the Universe'' and BCCP Fellow. SKG acknowledges support from Olle Engkvist Stiftelse (grant no.
232-0238). We have utilised the following Python packages for manipulating the simulation outputs and plotting results: \texttt{numpy} \citep{harris2020array}, matplotlib \citep{Hunter2007} and {\sc tools21cm}\footnote{\url{https://github.com/sambit-giri/tools21cm}} package \citep{Giri2020}.
\section*{Data Availability}
The data underlying this article will be shared upon reasonable request
to the corresponding author.



\bibliographystyle{mnras}
\bibliography{example} 




\bsp	
\label{lastpage}
\end{document}